\documentclass[12pt]{iopart}

\usepackage{iopams}  

\usepackage[utf8]{inputenc}
\usepackage{hyperref}

\def\be{\begin{equation}}
\def\ee{\end{equation}}

\begin{document}

\title{On slowly rotating black holes and nonlinear electrodynamics}

\author{Claus L\"ammerzahl}
\address{ZARM, Universit\"at Bremen, \\Am Fallturm, 28359 Bremen Germany}
\ead{laemmerzahl@zarm.uni-bremen.de}

\author{Marco Maceda}
\address{Departamento de F\'{\i}sica, Universidad Aut\'onoma Metropolitana-Iztapalapa,\\A.P. 55-534, CDMX 09340, M\'exico.}
\ead{mmac@xanum.uam.mx}

\author{Alfredo Mac\'{\i}as}
\address{Departamento de F\'{\i}sica, Universidad Aut\'onoma Metropolitana--Iztapalapa,\\A.P. 55-534, CDMX 09340, M\'exico.}
\ead{amac@xanum.uam.mx}

\vspace{10pt}
\begin{indented}
\item[]June 2018
\end{indented}

\begin{abstract}
We discuss the solution to Einstein's equations for a Lense-Thirring inspired metric describing a slowly rotating black hole coupled to nonlinear electrodynamics. We show that different schemes of rotation for the black hole exist; they depend on a parameter $\gamma$ defining the dependence of the metric on the polar angle. The fulfilment of the complete set of gravitational field equations and conservation laws implies constraints on this parameter and the metric functions. The vanishing of $\gamma$ provides the Lense-Thirring line element associated to any non-linear electrodynamics; the Kerr-Newman metric for slow rotation arises when $\gamma$ is not vanishing, a feature that emphasises the unique role played by Maxwell's electrodynamics. 
\end{abstract}

%
\vspace{2pc}

\section{Introduction}
\label{secc:1}

Black holes have been a subject of great interest for the last decades due to its conceptual and physical implications. In recent years an increasing number of related observational data have provided us with the opportunity to analyse the structure of such an object. In particular, a lot of effort has been given to the study of the black hole at the centre of our galaxy, Sagittarius A*, and several of its physical parameters have been estimated using the effects on the orbits of its neighbouring stars.

As theoretical constructions, black hole solutions are an appropriate playground to test theories that modify General Relativity at high energy scales. Some approaches to a quantum theory of gravity, for example, explore the consequences of a discrete structure of spacetime and this changes in a non-trivial way the appearance and nature of horizons; other models analyse the implications of additional scalar or tensor fields coupled to gravity. Spacetimes describing black holes in higher dimensions are also useful tools in string and braneworld theories.

Since black holes are working laboratories where high energies and nonlinearities are involved, the behaviour of charged particles around them may be described by non-linear electrodynamics (NED); hence it is worthwhile to address the question of the gravitational field coupled to a NED. This issue has been discussed in the literature for many years and the Einstein-Born-Infeld (EBI) spacetime~\cite{Born:1934ji,Born:1934gh} is one of the most used models to describe this behavior; it has also been extensively considered in other contexts~\cite{Wiltshire:1988uq,Rasheed:1997ns,Tseytlin:1999dj,Tamaki:2000ec,Gibbons:2001gy,Breton:2003tk,Aiello:2004rz,Cataldo:1999wr,Fernando:2003tz,Cai:2004eh,Dey:2004yt}. For specific values of its parameters, it can remove the curvature singularity that is often present in other models such as Reissner--Nordström where Maxwell's electrodynamics is involved.

Nowadays, there exist a revival of interest in non--linear electrodynamics since in the effective theory arising from superstrings a generalised Born--Infeld action naturally occurs as the leading part of the effective string action\cite{Fradkin:1985qd,Tseytlin:1986ti,Leigh:1989jq,DeFosse:2001mk}. Besides, much attention has been given to the interpretation of the solutions to the Born--Infeld equations as states of D--branes~\cite{Gibbons:1997xz}.

Additionally, quantum electrodynamical vacuum corrections to the Maxwell--Lorentz theory can be accounted for by an effective non--linear theory derived by Euler and Heisenberg \cite{Heisenberg:1935qt,Weinberg:2005a}. The vacuum is treated as a specific type of medium, the polarizability and magnetizability properties of which are determined by the clouds of virtual charges surrounding the real currents and charges \cite{Obukhov:2002xa}. Recently, Bordin et al. \cite{Brodin:2001zz} proposed a possible direct measurement of the Euler--Heisenberg effect. This theory is a valid physical theory \cite{Boillat:1970gw}, and it is the low field limit of the Born--Infeld one \cite{Hehl:2003}.

EBI spacetime describes, however, a static source, the EBIon~\cite{Demianski:1986wx}. Inspired in the concept of geon, a soliton-like solution where Maxwell's electromagnetic radiation is held together by its self-gravitating attraction, the EBIon is the corresponding configuration when dealing with BI electrodynamics~\cite{Breton:2002it}. Rotating black hole solutions are more difficult to obtain, as evidenced by the lapse of almost 50 years between the Schwarzschild and Kerr solutions. The search for black hole solutions with rotation has been an open problem since the publication of the Kerr solution \cite{Kerr:1963ud} and the Kerr--Newman solution \cite{Newman:1965my}. The Kerr--Newman solution is the most general stationary black hole solution to the Einstein-Maxwell system of field equations. This solution provides valuable insights into the other black hole solutions, in particular, the Kerr one. In the case of coupling to a NED, such as Born-Infeld electrodynamics, recent efforts focus on the behaviour to lowest order on the rotation parameter~\cite{CiriloLombardo:2006ph}. Many attempts to obtain rotating solutions have used the Newman--Janis algorithm \cite{Newman:1965tw}, nevertheless it does not work with nonlinear sources \cite{Rodrigues:2017tfm}, i.e., the resulting solution does not satisfy the whole set of field equations of the theory.

In a recent paper~\cite{Hendi:2014xia}, some solutions to the gravitational field equations coupled to NED are analysed in the case of a spacetime metric describing a slowly rotating black hole; the relevant dependence on the rotation parameter is encoded in the only non-vanishing off-diagonal metric coefficient $g_{t\phi}$. This coefficient involves a function of the polar angle, which when appropriately chosen, allows an exact solution, even though on physical grounds one may object to that choice since this does not correspond to a real physical situation.

In this work, we reconsider this problem. For this purpose, we use a Lense-Thirring inspired metric in the limit of slow rotation. We show then that contrary to~\cite{Hendi:2014xia}, the angular part of $g_{t\phi}$ allows a whole 1-parameter family of solutions depending on a parameter $\gamma$. From this family, a solution exists with the property to be symmetric with respect to the equatorial plane. The main consequences from this result are that the general solution for a given NED corresponds to $\gamma = 0$ and the unique character of Maxwell's electrodynamics arises when $\gamma=2$, providing us with an asymptotically flat solution. 

We organise the paper as follows: First, in Sec.~\ref{secc:2}, we introduce the Lense-Thirring inspired metric and give the main ingredients of the model. We then analyse the associated conservations laws and the gravitational field equations in Sec.~\ref{secc:3}; there we examine the exact fulfilment of this set of equations, determine the values of the parameter $\gamma$ and write the associated Lense-Thirring inspired metrics. From these results, we discuss the relationship between the metric coefficients and the parameter $\gamma$ in Sec.~\ref{secc:4}; we show that indeed the values $0$ or $2$ for $\gamma$ cover all the possible NED. We finally end with some comments and perspectives in the Conclusions.

\section{Metric and electromagnetic tensor}
\label{secc:2}

After the publication of the Schwarzschild solution describing the gravitational field of a static mass, Lense and Thirring found a metric representing the external gravitational field of a slowly rotating mass~\cite{Lense:1918}. The Lense-Thirring (LT) spacetime has the following length element
\be
ds^2 = - \left( 1- \frac {2m}r \right) dt^2 + \left( 1- \frac {2m}r \right)^{-1} dr^2 + r^2 d\theta^2 + r^2 \sin^2 \theta \left( d\phi - \frac {2am}r dt \right)^2,
\ee
and it is a solution to the vacuum gravitational field equations if terms quadratic on the angular parameter $a$ are neglected (slow-rotation). In this last situation, we obtain the simplified form
\be
ds^2 = - \left( 1- \frac {2m}r \right) dt^2 + \left( 1- \frac {2m}r \right)^{-1} dr^2 + r^2 d\theta^2 + r^2 \sin^2 \theta d\phi^2 - \frac {4am\sin^2 \theta}r d\phi dt. 
\ee

Inspired by the LT metric, we consider the following LT-like line element
\begin{eqnarray}
ds^2 &=& -G(r) dt^2 + \frac 1{G(r)} dr^2 + r^2 d\theta^2 + r^2 \sin^2 \theta [d\phi + a F(r) K(\theta) dt]^2
\nonumber \\[4pt]
&=& -G(r) dt^2 + \frac 1{G(r)} dr^2 + 2a F(r) K(\theta) dt d\phi + r^2 (d\theta^2 + \sin^2 \theta d\phi^2),
\end{eqnarray}
as describing the gravitational field of a slowly rotating source coupled to a NED; here $a$ is the angular momentum per unit mass that we assume to be small in the following. The field equations and conservation laws will determine the functions $G(r), F(r)$ and $K(\theta)$ of the metric.

The physical content of the NED is obtained from the electromagnetic potential
\be
A = h(r) [dt + a K(\theta) d\phi].
\ee
In consequence, the electromagnetic tensor has the following expression
\be
F = dA = h_{,r} dr \wedge dt + a h_{,r} K dr \wedge d\phi + a h K_{,\theta} d\theta\wedge d\phi,
\ee
where ${}_{,r}$ and ${}_{,\theta}$ mean differentiation with respect to $r$ and $\theta$ respectively. The non-vanishing components $F_{\mu\nu}$ are thus
\be
F_{10} = h_{,r}, \qquad F_{13} = a h_{,r} K \qquad F_{23} = a h K_{,\theta},
\ee
and from them it follows that the only non-vanishing contravariant components are
\be
F^{01} = h_{,r}, \qquad F^{13} = \frac {a (F + G) h_{,r} K}{r^2 \sin^2 \theta}, \qquad F^{23} = \frac {a h K_{,\theta}}{r^4 \sin^2 \theta},
\ee
neglecting terms quadratic on $a$. Here we have used the inverse metric
\be
g^{\mu\nu} = \left (
\begin{array}{cccc}
-1/G & 0 & 0 & aFK/G r^2\sin^2 \theta
\\[4pt]
0 & G & 0 & 0
\\[4pt]
0 & 0 & 1/r^2 & 0
\\[4pt]
aFK/G r^2\sin^2 \theta & 0 & 0 & 1/r^2 \sin^2 \theta
\end{array}
\right),
\ee
valid up to linear terms on $a$. We also have $\sqrt{-g} = r^2 \sin \theta$ in this approximation, and a straightforward calculation shows that the first electromagnetic invariant is
\be
{\cal F} := F_{\mu\nu} F^{\mu\nu} = -2 h_{,r}^2.
\ee

\section{Conservation laws and gravitational field equations}
\label{secc:3}

Since we assume the angular parameter $a$ to be small, we consider electromagnetic Lagrangians for a NED depending solely on the invariant $\cal F$. For a Lagrangian  
\be
L = L( {\cal F}), \qquad {\cal F} = F_{\mu\nu} F^{\mu\nu} = - 2 h_{,r}^2 = -2 E^2,
\ee
the conservation laws for the electromagnetic field are
\be
\partial_\mu (\sqrt{-g} L_{\cal F} F^{\mu\nu}) = 0\, ,
\label{CL1}
\ee
where $L_{\cal F} := dL({\cal F})/d{\cal F}$. These equations are equivalent to
\be
\partial_\mu (\sqrt{-g} F^{\mu\nu}) = - L_{\cal F}^{-1} (\partial_\mu L_{\cal F}) \sqrt{-g} F^{\mu\nu}\, ,
\label{CL1a}
\ee
and from them, we obtain the following set of equations
\begin{eqnarray}
&&E_{,r} + \frac {2E}r \frac 1{1 - 4 E^2 (L_{\cal F}^{-1} L_{\cal FF})} = 0,
\nonumber \\[4pt]
&&(F + G) [1 - 4 E^2 (L_{\cal F}^{-1} L_{\cal FF})] E_{,r} + (F + G)_{,r} E + \frac h{r^2} \frac {K_{,\theta\theta} - \cot \theta K_{,\theta}}K = 0,
\label{conslawsgned}
\end{eqnarray}
where $L_{\cal FF} := d^2 L({\cal F}) /d{\cal F}^2$. We notice that in order to eliminate any dependence on the polar angle $\theta$ in the last equation, the function $K(\theta)$ must satisfy the following ordinary differential equation
\be
K_{,\theta\theta} - \cot \theta K_{,\theta} + \gamma K = 0,
\label{equ4k}
\ee
where $\gamma$ is an arbitrary constant. The general solution to Eq.~(\ref{equ4k}) is
\begin{eqnarray}
K(\theta) &=& c_1 \, {}_2F_1\left( -\frac 14 - \frac 14 \sqrt{1 + 4\gamma}, -\frac 14 + \frac 14 \sqrt{1 + 4\gamma}; \frac 12; \cos^2 \theta \right)
\nonumber \\[4pt]
&&+ c_2 \cos \theta \, {}_2F_1 \left( \frac 14 - \frac 14 \sqrt{1 + 4\gamma}, \frac 14 + \frac 14 \sqrt{1 + 4\gamma}; \frac 12; \cos^2 \theta \right),
\label{sol4k}
\end{eqnarray}
where ${}_2F_1(a,b; c; z)$ is the Gaussian hypergeometric function and $c_1,c_2$ are constants. Two particular cases are worthwhile to mention: when $\gamma = 0$ we have
\be
K(\theta) = c_1 + c_2 \cos \theta.
\label{case1}
\ee
Solutions with this form have been discussed already in~\cite{Hendi:2014xia}; if the regularity condition $K(0)=K(\pi) =0$ is imposed, then this value for $\gamma$ is prohibited. Another case is $\gamma = 2$ leading to
\be
K(\theta) = c_1 \sin^2 \theta + c_2 g(\cos \theta),
\label{case2}
\ee
where $g$ is a complicated function involving powers and logarithms of its argument; this solution is related to Kerr-Newman spacetime as we shall see later. It should be remarked that the general solution in Eq.~(\ref{sol4k}) is not invariant under the transformation $\theta \to \pi - \theta$ if $c_2 \neq 0$; this means that by setting $c_2 = 0$ we have solutions symmetric with respect to the equatorial plane. We shall see later that the values $\gamma = 0$ and $\gamma = 2$ correspond to arbitrary NED and Maxwell electrodynamics respectively.

If $\gamma$ is an even number, Eq.~(\ref{sol4k}) is rewritten in terms of Jacobi polynomials $P^{(\alpha, \beta)}_n(z)$ as
\be
K(\theta) = \tilde c_1 P^{(-\frac 12, -1)}_n (1 - 2\cos^2\theta) + \tilde c_2 P^{(-\frac 12, 0)}_m (1 - 2\cos^2\theta),
\ee 
with $2n(2n-1) = \gamma = 2m(2m+1)$ and $\tilde c_1, \tilde c_2$ constants. Angular distributions depending on Jacobi polynomials are not so often found in the literature; most frequently we encounter Legendre polynomials. It is straightforward to check that the values $n=0,1$ and $m=0$ reduce the Jacobi polynomials to Legendre polynomials in the above expression; they correspond to the values $\gamma$ equal to zero and two.

Once the function $K$ is fixed by its differential equation, we can find an expression for the sum $F + G$ in Eqs.~(\ref{conslawsgned}); elimination of $E_{,r}$ from these equations leads to 
\be
(F + G)_{,r} - \frac 2r (F + G) = \gamma \frac {h}{r^2 E}.
\label{equfandg}
\ee
Therefore, we obtain by a straightforward calculation that
\be
F + G = r^2 \left[ \gamma \int^r \frac {h(s)}{E(s)} \frac {ds}{s^4} + c \right], \qquad c = const.
\label{fplusg}
\ee
This expression links the metric functions $F$ and $G$ and plays the role of an equation of motion coming from the conservation laws. It is worth to notice that if $c=0$ and $E = q/r^2$, or equivalently $h = -q/r$, then we obtain the simple relation
\be
F + G = -\gamma r^2 \int^r \frac {ds}{s^3} = \frac \gamma2.
\label{fg4maxwell}
\ee

Let us now look to the gravitational field equations
\be
G_{\mu\nu} + \Lambda g_{\mu\nu} = \frac 12 g_{\mu\nu} L({\cal F}) - 2 L_{\cal F} F_{\mu\lambda} F_\nu^\lambda,
\ee
where $\Lambda$ is the cosmological constant. First, we find the components of the Einstein tensor $G_{\mu\nu}$; they are
\begin{eqnarray}
&&G_{00} = -\frac Gr (-1 + G + G_{,r}), 
\nonumber \\[4pt] 
&&G_{11} = \frac 1{r^2 G} (-1 + G + G_{,r}),
\nonumber \\[4pt]
&&G_{22} = \frac 12 (r^2 G_{,r})_{,r}, 
\nonumber \\[4pt] 
&&G_{33} = \sin^2 \theta G_{22},
\nonumber \\[4pt]
&&G_{03} = \frac {aK}{2r^2} [-r^2 GF_{,rr} - \frac {K_{,\theta\theta} - \cot \theta K_{,\theta}}K F + F (-2 + 2G + (r^2 G_{,r})_{,r}) ].
\end{eqnarray}
Using the differential equation for the function $K$, the component $G_{03}$ is simplified to
\be
G_{03} = \frac {aK}{2r^2} [-r^2 GF_{,rr} + F(\gamma -2 + 2G + (r^2 G_{,r})_{,r})].
\ee
In consequence, the relevant gravitational field equations are
\begin{eqnarray}
r G_{,r} + G - 1 +\Lambda r^2 + \frac {r^2}2 (-L + 2{\cal F} L_{\cal F}) = 0,
\nonumber \\[4pt]
r G_{,rr} + 2G_{,r} + 2\Lambda r + r (-L) = 0,
\nonumber \\[4pt]
r^2 G F_{,rr} - 2 \left( G - 1 + \frac \gamma2 \right) F + 2 Gr^2 {\cal F} L_{\cal F} = 0.
\label{fequsgned}
\end{eqnarray}
We can see that the second line in Eqs.~(\ref{fequsgned}) is a consequence of the first line by differentiation of the latter with respect to $r$; use of the relations
\begin{eqnarray}
{\cal F}_{,r} &=& \frac {8E^2}r \frac 1{1 - 4 E^2 (L_{\cal F}^{-1} L_{\cal FF})},
\nonumber \\[4pt]
(-L + 2{\cal F} L_{\cal F} )_{,{\cal F}} &=& L_{\cal F} (1 + 2 L_{\cal F}^{-1} L_{\cal FF}),
\nonumber \\[4pt]
1 - 4 E^2 (L_{\cal F}^{-1} L_{\cal FF}) &=& 1 + 2 L_{\cal F}^{-1} L_{\cal FF},
\end{eqnarray}
proves helpful for this purpose. On the other hand, the left hand side of the third line in Eqs.~(\ref{fequsgned}) can also be obtained by differentiation of Eq.~(\ref{equfandg}) with respect to $r$ and using afterwards the first and second lines in Eqs.~(\ref{fequsgned}). The final outcome is then
\be
r^2 G F_{,rr} - 2 \left( G - 1 + \frac \gamma2 \right) F + 2 Gr^2 {\cal F} L_{\cal F} = (2 - \gamma) F + G \left( 2 + \gamma - \gamma \frac {hE_{,r}}{E^2} \right).
\ee
For this result to be consistent with the third gravitational field equation, we must have
\be
(2 - \gamma) F + G \left( 2 + \gamma - \gamma \frac {hE_{,r}}{E^2} \right) = 0.
\label{constraint}
\ee
We obtain then a constraint equation for the model involving the metric functions $F$ and $G$. An equivalent writing of this condition is
\be
\gamma \left( 2 r^2 \int^r \frac {h(s)}{E(s)} \frac {ds}{s^4} + G - F - \frac {hE_{,r}}{E^2} G \right) + 2c r^2 = 0,
\label{constraintb}
\ee
obtained by means of Eq.~(\ref{fplusg}). The existence of this constraint is due to the {\it fulfillment of the complete set} of gravitational field equations and conservation laws in nonlinear electrodynamics, Eqs.~(\ref{fequsgned}) and~(\ref{CL1}) respectively. 

There are at least two particular cases when the constraint in Eq.~(\ref{constraint}) holds:
\begin{enumerate}
\item Set $\gamma = 0, c = 0$, then $F+ G = 0$ and we also take $K (\theta) = \cos \theta$. We have no restrictions on the electric field since Eqs.~(\ref{constraint}) or~(\ref{constraintb}) are identically satisfied and thus the NED can be arbitrary. The metric function $G$ is obtained from the first line in Eqs.~(\ref{fequsgned}) as
\be
G(r) = 1 - \frac {2M}r - \frac \Lambda3 r^2 - \frac 1{2r} \int^r (-L + 2{\cal F} L_{\cal F}) s^2 ds,
\label{g1}
\ee
once the Lagrangian of the NED is given and the electric field is found as a function of $r$. Let us set the cosmological constant to vanish. Even though acceptable as a solution to the field equations, the fact that $F + G = 0$ implies that the function $F$ goes to $-1$ as $r$ tends to infinity. If we want an asymptotically Kerr-Newman spacetime such behaviour for $F$ is not admissible. For this solution, the Lense-Thirring inspired line element is
\begin{eqnarray}
ds^2 &=& -G(r) dt^2 + \frac 1{G(r)} dr^2 + r^2 d\theta^2 + r^2 \sin^2 \theta [d\phi - a G(r) \cos\theta dt]^2.
\label{lt1}
\end{eqnarray}
To this class belong the solutions discussed by Hendi and Allahverdizadeh \cite{Hendi:2014xia}. For example, if we consider the exponential NED defined by the Lagrangian
\be
L_{ENED} = \beta^2 (e^{-{\cal F}/\beta^2} - 1),
\ee 
with ${\cal F} = -2 E^2 = -2 q^2 \exp[-LambertW(4q^2/(\beta^2 r^4))]/r^4$, then 
\be
-L + 2 {\cal F} L_{\cal F} = \beta^2 -\beta^2 e^{2E^2/\beta^2} + 4E^2 e^{2E^2/\beta^2}. 
\ee
Continuous use of the relation 
\be
r^2 e^{2E^2/\beta^2} = \frac {2q}{\beta \sqrt{LambertW(4q^2/(\beta^2 r^4))}},
\ee
in the integrals of Eq.~(\ref{g1}) allows us to recover the exponential NED metric. Similar considerations apply to other NED.

It is clear that the line element Eq.~(\ref{lt1}) lacks equatorial symmetry; we can mend this last issue if we chose $K (\theta) = 1$ since this solution is allowed (see Eq.~(\ref{case1})), but this choice also lacks physical significance. 

\item Set $\gamma = 2, c = 0$, then we choose $K(\theta) = \sin^2 \theta$. From Eq.~(\ref{constraint}) the electric field is constrained to satisfy
\be
2 E^2 = h E_{,r}, \qquad E = h_{,r}.
\ee
with solution
\be
h = - \frac qr, \qquad E = \frac q{r^2},
\ee
associated to classical Maxwell electrodynamics. Indeed, from the equation for $E$ in the first line in Eqs.~(\ref{conslawsgned}), we have that $L_{\cal FF} = 0$; therefore, the Lagrangian must be linear on ${\cal F}$. As indicated previously in Eq.~(\ref{fg4maxwell}), we obtain
\be
F + G = 2 r^2 \int^r \frac {h(s)}{E(s)} \frac {ds}{s^4} = 2r^2 \cdot \frac 1{2r^2} = 1.
\ee
Furthermore, from the field equations we have
\be
G(r) = 1 - \frac {2M}r - \frac \Lambda3 r^2 + \frac {q^2}{r^2}.
\label{g2}
\ee
The Lense-Thirring inspired line element is
\be
ds^2 = -G(r) dt^2 + \frac 1{G(r)} dr^2 + r^2 d\theta^2 + r^2 \sin^2 \theta (d\phi + a [1 - G(r)] \cos^2\theta dt)^2.
\ee
If $\Lambda = 0$, $G \to 1$ as $r\to \infty$, we have $F = 2M/r + \Lambda r^2/3 -q^2/r^2 \to 0$ asymptotically. In this scenario, we have then an asymptotically flat spacetime, and we recover the Kerr-Newman metric if we neglect quadratic contributions on the rotation parameter $a$. 
\end{enumerate}

Had we considered $c \neq 0$ in the previous cases, then we would have the following results: When $\gamma = 0$, Eqs.~(\ref{fplusg}) and~(\ref{constraint}) imply 
\be
F + G = c r^2, \qquad F + G = 0,
\ee
and hence we obtain $c=0$ and Eq.~(\ref{g1}) as before. On the other hand, when $\gamma = 2$, Eqs.~(\ref{fplusg}) and~(\ref{constraint}) imply 
\be
F + G = r^2 \left( \int^r \frac {h(s)}{E(s)} \frac {ds}{s^4} + c \right) , \qquad G\left( 2 - \frac {h E_{,r}}{E^2} \right) = 0,
\ee
Since $G \neq 0$, we have $h=-q/r, E = q/r^2$ and $F + G = 1 + cr^2$. The Lense-Thirring inspired line element is
\be
ds^2 = -G(r) dt^2 + \frac 1{G(r)} dr^2 + r^2 d\theta^2 + r^2 \sin^2 \theta [d\phi + a F(r) \cos^2\theta dt]^2.
\ee
with
\be
F(r) = 1 + cr^2 - G = \frac {2M}r + \left( c + \frac \Lambda3 \right)r^2 - \frac {q^2}{r^2}.
\ee
using Eq.~(\ref{g2}). The constant $c$ may be used to eliminate the presence of the cosmological constant into the metric function $F$. As before, if $\Lambda = 0 = c$, then $F \to 0$ asymptotically; if $\Lambda \ne 0 = c$ we have an asymptotically Kerr-Newman-ADS black hole.

\section{Determination of metric functions $G, F$ and parameter $\gamma$}
\label{secc:4}

As seen from the previous section, the values zero and two for $\gamma$ seem to play a unique role in the solutions. In this section, we explore the possibility that this parameter may take other values.  From Eq.~(\ref{fplusg}) we write $F$ in terms of $G$, and then use this in Eq.~(\ref{constraint}); we obtain thus
\be
\gamma \left[  2r^2 \int^r \frac {h(s)}{E(s)} \frac {ds}{s^4} + \left( 2 - \frac {h E_{,r}}{E^2} \right) G - \gamma r^2 \int^r \frac {h(s)}{E(s)} \frac {ds}{s^4}\right] + (2 - \gamma) cr^2 = 0.
\ee
The above expression is satisfied when $\gamma$ is zero or two as previously discussed. Let us assume that a different value for $\gamma$ is permisible, then it should happen that
\be
\left( 2 - \frac {h E_{,r}}{E^2} \right) G = \tilde \gamma r^2 \int^r \frac {h(s)}{E(s)} \frac {ds}{s^4} + \delta r^2,
\ee
where $\tilde \gamma, \delta$ are constants different from zero. Now, for large distances the left hand side in this expression vanishes if we demand an asymptotically Kerr-Newman spacetime since in this case $h \sim 1/r, E \sim -1/r^2, E_{,r} \sim 2/r^3$ and $G \sim 1$. In consequence
\be
\lim_{r\to \infty} r^2 \int^r \frac {h(s)}{E(s)} \frac {ds}{s^4} + \frac \delta{\tilde \gamma} r^2 = 0,
\ee
since $\tilde \gamma \neq 0$. The above constraint means that
\be
\int^r \frac {h(s)}{E(s)} \frac {ds}{s^4} = - \frac \delta{\tilde \gamma} + \frac a{r^\alpha} + \dots, \qquad a = const.,
\ee
with $\alpha \geq 3$ at least. Differentiation with respect to $r$ allows us to write then the following equation for $h$ (recall that $E = h_{,r}$)
\be
\frac {h_{,r}}h = - \frac 1{a\alpha} r^{\alpha - 3},
\ee
to leading order; its solution is
$$
h = h_0 \exp \left[ -\frac {r^{\alpha - 2}}{a \alpha(\alpha - 2)} \right].
$$
Since $\alpha \geq 3$, this exponentially decreasing behavior of $h$ for large $r$ contradicts the assumed Kerr-Newman behavior. It follows that only the values zero and two for $\gamma$ are permissible.

\section{Conclusions}

We have analysed the solution to the gravitational field equations for a slowly rotating black hole in the presence of a NED; for that purpose, we considered a Lense-Thirring inspired metric that describes a weak gravitational field in the slow rotating case. Given the fact that the Newman-Janis algorithm fails to generate solutions to the field equations for arbitrary rotation in the presence of nonlinear electrodynamics, the analysis of this situation represents a step in this direction.   

We have shown that to solve the {\it whole} set of field equations and conservation laws in exact form we need to introduce a parameter $\gamma$ that decouples the dependence on the polar angle in one of the conservation equations. Different values of this parameter lead to different functions $K(\theta)$ in the metric coefficient $g_{t\phi}$ of the rotating black hole metric. We showed that $\gamma$ equal zero or two leads to different forms of the Lense-Thirring inspired metric. The case $\gamma = 2$ is particularly important since it fixes the NED to be linear; in this case, for slow rotation, the Lense-Thirring inspired metric gives the expected behaviour of a slowly rotating Kerr-Newman black hole. We explored as well the possibility of values for $\gamma$ different from zero and two, and we concluded that these are the only allowed values.

In \cite{Hendi:2015ixa} dilatonic black hole solutions were investigated using an Ansatz for the gauge potential that involves two different functions, $h(r)$ and $C(r)$, of the radial coordinate $r$; these two functions are never equal, $h(r)$ is obtained by integration of Eq.~(15) and $C(r)$ is given in Eq~(30) of that paper. In our work, there is only one function $h(r)$ for the description of the gauge potential. and we are thus closer to the situation of the Kerr-Newman metric, where there is only one function involved. It would be interesting nevertheless to extend our results for NED to the case where two different functions appear in the gauge potential. 

Finally, we should mention that propagation of light rays may also be analysed using the so--called Pleba\'nski's pseudo--metric $\gamma^{\mu\nu}$~\cite{Plebanski:1970,Novello:1999pg} defined as follows
\begin{equation}
\gamma^{\mu\nu} = p g^{\mu\nu} + r T^{\mu\nu},
\end{equation}
with $g^{\mu\nu}$ the metric and $T^{\mu\nu}= - F^{\mu\rho}F^\nu{}_\rho + Fg^{\mu\nu}$ the energy--momentum tensor for Maxwell's electrodynamics; here $p$ and $q$ are arbitrary scalar fields. This pseudometric satisfies the relation $Det||\gamma^{\mu\nu}|| = Det||g^{\mu\nu}|| (p^2-r^2|F + {}^*F|^2)^2$ and its covariant components are
\begin{equation}
\gamma_{\mu\nu} = \frac{p g^{\mu\nu} - r T^{\mu\nu}}{p^2-r^2|F + {}^*F|^2}.
\end{equation}
The effective metric $\gamma_{\mu\nu}$ gives us complementary information about the trajectories followed by photons; this analysis will be reported elsewhere.

\ack

The authors would like to thank PD Dr. Volker Perlick for useful discussions on the subject. We also want to thank the anonymous referees for their valuable comments and suggestions. This work was supported by CONACyT Grant No. A1-S-31056. C.L. acknowledges support by the DFG Research Training Group 1620 \emph{Models of Gravity} and by the QUEST Center of Excellence. A.M. acknowledges support from DAAD.

\section*{References}


\providecommand{\newblock}{}

\end{document}